\documentclass[aps,preprint,10pt,onecolumn]{revtex4}%
\usepackage{amsfonts}
\usepackage{amsmath}
\usepackage{amssymb}
\usepackage{graphicx}%
\providecommand{\U}[1]{\protect\rule{.1in}{.1in}}
\begin{document}
\preprint{cond-mat 01/2008}
\title[$FeSe$]{Magnetism and exchange coupling in iron pnictides}
\author{J.J.Pulikkotil$^{1}$, L. Ke$^{2}$, M. van Schilfgaarde$^{2}$, T. Kotani$^{2}$,
and V.P.Antropov$^{1}$}
\affiliation{$^{1}$ Condensed Matter Physics, Ames Laboratory, IA 50011}
\affiliation{$^{2}$ School of Materials, Arizona State University}
\pacs{PACS number}

\begin{abstract}
Using linear-response density-functional theory, we obtain the magnetic
interactions in the several iron pnictides. The ground state has been found to
be non-collinear in FeSe, with a large continuum of nearly degenerate states
lying very close to the magnetic \textquotedblleft striped\textquotedblright%
\ structure. The presence of non-collinearity also seems to be a generic
feature of iron pnictides when the Fe moment is small. At small R$_{Fe-Se}$
the system is itinerant: strong frustration give rise to excess of spin
entropy, long ranged interactions create incommensurate orderings and strong
biquadratic (or ring) couplings violate the applicability of Heisenberg model.
There is a smooth transition to more localized behavior as R$_{Fe-Se}$
increases: stable magnetic orbital order develops which favor long range AFM
stripe ordering with strongly anisotropic in-plane exchange couplings. The
stabilization of the stripe magnetic order is accompanied by the inversion of
the exchange coupling.

\end{abstract}
\eid{identifier}
\date{\today}
\maketitle

Accumulated experimental studies indicate that the recently discovered iron
oxyarsenide family (Fe-As-La-O) and iron chalcogenide (Fe-Se)
superconductors\cite{EXPER,STRFESE} similar itinerant antiferromagnetic
metallic systems. This fact distinguishes these systems from traditional
high-$T_{c}$ superconductors where the local-moment picture and the importance
of strongly correlated electrons is generally accepted. While the physics of
these systems can be dramatically different they share an important common
feature: antiferromagnetic (AFM) interactions on a (nearly) 2D lattice. One
can hope that these materials have similar physics and analysis of new
metallic systems will shed some light on a nature of magnetic fluctuations in
both type of superconductors.

Description of magnetism in itinerant magnets is rather complicated, owing to
the lack of a commonly accepted rigid spin picture and the disappearance of
adiabaticity. The trivial use of localized models leads to oversimplification
of the physics of itinerant magnets. As we show here, the degree of itineracy
is critical to understand the ground states and magnetic excitations in iron
pnictides. Owing to a strong competition between on-site Hund energy and Fe-As
bonding in iron pnictides\cite{Mirbt,PHILIPS,Kir}, we adopt the local density
approximation (LDA) as we believe it to be the most appropriate readily
accessible method to describe these systems\cite{NOTE}.

We analyze the magnetic properties of iron pnictides, using a variety of
implementations of the LSDA, including FLAPW, FPLMTO, LMTO-ASA. The latter
includes calculations of non-collinear configurations following the original
prescription\cite{NONCOL}. Full-potential results are in excellent agreement
with each other, and (where they can be compared) very similar to those
obtained in Ref.\cite{SinghFeSe}. The ASA results are also in reasonably good
agreement with them. For the linear response calculations we adopted a Green's
function formalism within the ASA~\cite{EX1,Sam}. This technique has been
extensively tested and applied to many different metallic and insulating
systems\cite{EX1,LINRESP}. A similar method was recently applied to
LaFeAsO\cite{Savr1}. We have taken care to converge the calculations of
parameters reported, e.g. using a 16$\times$16$\times$12 $k$-mesh and 24
energy points for the contour integration in the complex plane.

We begin our analysis with the low-temperature phase of the FeSe\cite{STRFESE}%
. At low temperature FeSe assumes an orthorhombic structure\cite{STRFESE},
space group \emph{Cmma}, with distortions of FeSe slabs similar to those
observed in FeAs layers in the iron oxyarsenide family. Experimental lattice
parameters were employed\cite{STRFESE} with Fe at (4b) and As at (4g) Wyckoff
positions. In both FeSe and LaFeAsO, there is an internal parameter $z$ not
fixed by symmetry, corresponding to the height of the chalcogen above the
(possibly distorted) square of Fe atoms lying in the $xy$ plane. $z$ fixes the
Fe-Se distance R$_{Fe-Se}$, but we treat it here as a parameter because of the
ambiguity in determining it even while the magnetic properties depend strongly
on R$_{Fe-Se}$\cite{Mirbt}. The similar dependencies of magnetic moment on
volume have been observed in other iron based magnets (see Ref.\cite{Nitr} for
the review).

Our initial magnetic structure was obtained by relaxing a 32 Fe atom supercell
(with $z$=0.25). Spins were initially put in random (noncollinear)
orientations and their orientations relaxed along with the spin and charge
densities, as described in Ref.\cite{SDPRB}. In this way several magnetic
structures have been identified (Fig.1). The self-consistent (minimum-energy)
\textit{collinear} solution was found to be an AFM configuration in the
striped geometry, with a Fe moment $M\sim0.85\mu_{B}.$ As Fig.~1 shows, $M$ is
very sensitive to R$_{Fe-Se}$. The N{\'{e}}el state is the least stable among
the structures indicated in Fig.~1: it is $\sim$1.0 mRy/Fe less stable than
the striped structure (and nonmagnetic at the LSDA equilibrium geometry). All
structures exhibit similar dependence on $M($R$_{Fe\text{-}Se})$. At very
small moments ($M${}$<${}$0.2\mu_{B}$) orderings $(b)$,$(c)$,$(d)$ in Fig.~1
have lower energy then the striped configuration $(a)$. For $M${}$>$%
0.95$\mu_{B}$ (large R$_{Fe\text{-}Se}$) the dimer structure (Fig.$(1e)$) has
lowest energy. Such sequence of ground states clearly indicate the
significance of couplings beyond first two nearest neighbors.

The following picture emerges from studies of the pairwise exchange coupling
parameters $J_{ij}$ and non-magnetic susceptibility $\chi_{ij}$ in different
structures, as a function of R$_{Fe\text{-}Se}$. We denote $\chi_{00}$ as the
on-site susceptibility and $\chi_{01}$ and $\chi_{02}$ that of the two nearest
neighbors (NN). Much of the behavior we observe depends on these three
quantities. Define $\eta=\chi_{02}/\chi_{01}$ as the \textquotedblleft
magnetic frustration\textquotedblright\ parameter (see below). For small $M$
(small R$_{Fe-Se}$), $\chi_{00}$ is rather small ($\sim$14-15 Ry$^{-1}$).
Noting that the Stoner parameter $I$ is $I${}$\approx$0.068 Ry for Fe, the
local moment criterion $I\chi_{00}>1$ is marginally, or not quite satisfied
(see Ref.\cite{Sam}). Thus the system already at this stage can be classified
as a marginally itinerant system: that is the magnetic moment depends on the
environment for its stability as opposed to localized behavior, where the
local moment is large and weakly dependent on environment. We find both
$\chi_{01}$ and $\chi_{02}$ are large and negative. So, the criterion for AFM
pair formation $I(\chi_{00}-\chi_{01})>1$ along (1,0), (0,1) and (1,1) is
fulfilled while the criterion of FM pair formation $I(\chi_{00}+\chi_{01})<1$
is not. FM fluctuations are thus strongly suppressed while AFM interactions
are frustrated, because both $\chi_{01}<0$ and $\chi_{02}<0$, with $\eta
\simeq0.4$. Increasing R$_{Fe-Se}$ (or volume) from a small value, a magnetic
moment appears as a consequence of the sum of exchange fields from all sites,
supporting the itinerant nature of the ground state there. This is
characterized by $\alpha=[\chi(q\text{=}0)/\chi_{00}-1]$, which differentiates
localized and itinerant systems\cite{LWA}. For small $M$, $\alpha${}$\sim$0.3.
As R$_{Fe\text{-}Se}$ increases to the magnetic instability point $I\chi=1$,
$I\chi\left(  \mathbf{q}\right)  $ approaches 1 in the small region of wave
vectors corresponding to striped and dimer orderings. At low moments AFM
states can be created and destroyed by arbitrarily weak interactions.

As R$_{Fe\text{-}Se}$ increases further, $M$ increases smoothly and $\eta$
decreases, reaching $\simeq$0.2 for large $M$. Thus the magnetic state evolves
to a less itinerant and less frustrated condition and even FM fluctuations can
exist. $\alpha$ decreases smoothly with $M$, but the magnetic behavior
continues to be borderline between itinerant and localized for intermediate
R$_{Fe-Se}$, and the frustrated nature of a \textit{weakly} magnetic state is
conserved for all reasonable distances R$_{Fe\text{-}Se}$. Only in the
strongly magnetic state is the situation drastically changed.

In Table 1 we present both longitudinal and transverse components of the
pairwise magnetic \textquotedblleft exchange\textquotedblright\ parameters
$J_{ij}$ for the striped structure with $M=0.9\mu_{B}$ (near the theoretically
optimized LSDA value). Both components of the exchange matrix $J$ have similar
amplitudes, indicating that both longitudinal and transverse spin fluctuations
must be taken into account. The NN coupling between atoms with parallel spins
(along {\emph{x}}, $\mathbf{\tau}$=1/2,0,0) is negative (\textquotedblleft
frustrated\textquotedblright) while the exchange between pairs with
antiparallel spins (along \emph{y}, $\mathbf{\tau}$=(0,1/2,0)) is positive
(also for $\mathbf{\tau}$=(1/2,1/2,0)). For more distant (1-5 to 1-9) in-plane
neighbors all parameters have similar amplitudes with different signs. While
two of the NN parameters indicate a stable pair configuration, Fig.~2 and the
Table show the presence of rather unstable FM pair alignments in plane, and
weak interaction along the $z$-axis (see $J_{001}$ in the Table 1), making
this system very nearly a 2D AFM. However, even such weak interplanar coupling
may lead to 3D long-range order and finite temperature transition. The insert
of Fig.~3 shows how the Fe moment in the striped structure depends on
anisotropy of the exchange coupling of NN atoms. The origin of this anisotropy
is as follows. For the N{\'{e}}el structure both $J_{1x}=J_{100}$ and
$J_{1y}=J_{010}$ are positive and equal (reflecting symmetry) while
$J_{2}=J_{110}$ is negative, reflecting instability of FM alignment along
(1,1). For the striped structure at small R$_{Fe-Se}$ (\textquotedblleft
bad\textquotedblright\ nesting conditions) the system still reflects the
frustrated $J_{1}$-$J_{2}$-$J_{3}$ model-like behavior and our $J_{1x}$ and
$J_{1y}$ have opposite signs with similar amplitudes (Fig.~2). As $M$
increases $J_{1}$ becomes anisotropic, reflecting a broken symmetry induced by
magnetic analog of Jahn-Teller effect\cite{Larkin,Zaa}. For $M\sim{}1.5\mu
_{B}$ the anisotropy is very large and soon 'frustrated' exchange coupling
along FM line disappears and the effective sign of this interaction becomes
positive. Thus, large moment state does correspond to localized Heisenberg
model and is absolutely stable with no sign of antiferromagnetic frustrations.
At high temperature both $x$ and $y$ directions are equivalent, but as $T$
decreases the barrier between $x$ and $y$ orientations will prevent
fluctuations between them and the spin symmetry-breaking phase---with
attendant structural distortions--will be frozen in. Fig.~$(1c)$ depicts the
"rotator" structure --- a state intermediate between $x$- and $y$- aligned
striped phases. At $M${}$\sim1\mu_{B}$ the LSDA energy of this state is 0.8
mRy/Fe relative to the striped state.

The spin wave velocities are not formed merely by NN interactions in such a
system, and the contribution from distant Fe atoms (3-4 \AA ) with
antiparallel spins is significant, as expected. We found significant in-plane
and out-of-plane anisotropies of spin velocities. For instance v$_{x}/$%
v$_{y}=1.6$ for $M=1.07\mu_{B}.$ This anisotropy has not been determined
experimentally. The calculated energy gap in the spin wave spectrum due to
spin orbital coupling\cite{REVIEW} is $\Delta=0.15$ mRy. At larger
$\mathbf{q}$ we see deviations from the adiabatic magnon picture. However the
adiabaticity parameter\ \cite{LWA} for $M${}$\sim1\mu_{B}$ is $\sim0.2$; so we
believe the system is already less itinerant and more adiabatic, therefore
antiferromagnons can be defined not only at small $\mathbf{q}$. However
longitudinal fluctuations of similar strength are always present and should be
seen in experiment. These features of spin excitation spectra will be
reflected in the temperature dependence of the sublattice magnetization and
will modify the enhancement of the linear specific heat. Also at those
R$_{Fe-Se}$ where the spin fluctuations are very localized around a particular
$\mathbf{Q}$ one may expect weak temperature dependence of the susceptibility.

Because parameters $J_{ij}$ strongly depend on magnetic configuration (which
may not be a ground state) and on symmetry-breaking, it is crucial to analyze
the magnetic interactions for several magnetic orderings. The universal
picture that emerges is that first and second NN interactions (both transverse
and longitudinal) on the square Fe sublattice are AFM and very strong, with
their absolute values and the ratio $\eta$ depending on R$_{Fe-Se}$. The role
of longitudinal fluctuations is crucial at larger $M$. Even so the relatively
small more distant neighbor couplings (and biquadratic terms) are not
negligible and important for the appearance of non collinear fluctuations at
small moments as we now describe.

To check the stability of our collinear "ground" striped state and avoid
uncertainties connected with the long-wave approximation\cite{LWA}, we
performed spin dynamics simulations with small deviations between FM coupled
atoms keeping first or second NN ordering antiferromagnetic. These excitations
include j$_{1}$ and j$_{2}$ modes (Fig.1) and fictitious spin spiral (SS)
orderings\cite{NONCOL}. Our results demonstrate that for those R$_{Fe\text{-}%
Se}$ with $M${}$<1\mu_{B}$ (Fig.3) spirals (\textbf{Q} + (0,\emph{q},0)) which
rotate AFM planes against each other exist only in some limited spin phase
space around $\mathbf{Q}$ of the striped structure, clearly establishing the
existence of the magnetic short range order (MSRO). This MSRO is significant
and can exist far above the critical temperature. However, the phase volume of
such long-wave excitations is rather small. As R$_{Fe\text{-}Se}$ increases
the short ranged excitations with much larger volume of phase space become
available and it leads to stabilization of the different magnetic ground state
or changes the character of excitations.

The total energy of SS configurations which rotate FM spins along the stripes,
i.e. (\textbf{Q} + (\emph{q},0,0)) appears very different. Fig.3 shows the
energy is nearly independent of $\mathbf{q}$ for a significant region of
$\mathbf{q}$. Such a flat surface in the itinerant AFM usually is a precursor
of frustrated interactions. This is exactly the present case, as evident from
the negative value of $J_{1x}$ (Fig~2). Moreover SS calculations show that the
collinear striped structure ($q$=0), which has strong AFM coupling between
stripes and no staggered magnetization at finite $q$, is not the
minimum-energy state, although the stabilization of this particular SS is very
small ($\sim0.01$ mRy) with SS $q$ consistent with the amplitude of
$J_{3},J_{4}$ couplings. We also found that j$_{1}$ fluctuating mode (which
has staggered magnetization) has very low excitation energy, so states in this
fluctuating direction are practically degenerate and only starting at
15-20$^{0}$ do we see a small energy increase $\sim0.03$ mRy. At larger angles
the energy rapidly increases, so the \textquotedblleft
zig-zag\textquotedblright\ (45$^{0}$) state has energy $\sim0.3$ mRy/Fe. The
nematic j$_{2}$ mode has much higher energy at large moments (which quite
natural for the biquadratic or ring type of exchange coupling), while at small
moments it has lower energy than the j$_{1}$ mode, supporting the increase of
frustration parameter $\eta$. Both "zig-zag" and "rotator" structures are
indicators of the strength of biquadratic exchange (K) and its importance for
the phase formation and spin wave spectrum gap. It is unusual to have
biquadratic exchange as large as half of bilinear exchange parameter for the
system with relatively small values of effective spin value (1/2 or 1).
However, we obtained a similar behaviour in many other pnictides and believe
that this is a generic feature of these systems. This fact, in turn, may
create favorable conditions for the applicability of model, suggested in
Ref.\cite{Larkin}. The biggest difference with that model consists in the
origin of barrier energy. In our calculations this barrier exists already on
adiabatic energy surface as a result of complicated spin-spin interactions in
the ground state, providing even stronger support of conclusions of
Ref.\cite{Larkin}. In addition, the low energy of j$_{1}$ and corresponding
'nematic' type of spin ordering (j$_{2}$) can be responsible for the
disappearance of LRO in plane due to developing at finite temperatures
fluctuations between x and y stripes. This mechanism of magnetic transition
can be considered as an alternative to usual formation of two-dimensional
state which is widely used in high temperature superconductors study. Besides
being important for the magnetic transition, these fluctuations must appear at
the finite temperatures, may persist in the paramagnetic case and should be
observable in the experiment. These type of excitations are  very unusual:
they do not exist at low temperatures in the ordered state, but will appear
closer to the Neel temperature or above it.

Thus, our exchange coupling and SS calculations indicate the presence of two
main competing mechanisms: one is an instability (driving force) towards
formation of AFM order ("good" nesting conditions), and second frustration
related to the strong negative couplings of first and second NN on a square
lattice (with $\eta$ $\sim0.2-0.4$). While in the $J_{1}$-$J_{2}$
model\cite{Larkin} $\eta$ is the only frustration parameter, our calculations
reflect a more complex picture: the relative strength of all frustrations (as
any interatomic coupling) is also determined by the ``itineracy''
characterized by $\alpha$.

The latter mechanism gives rise to large spin fluctuations related to $J_{3}$
or $J_{4}$ pair interactions and is particularly important at small $M$
(itinerant limit) with strong MSRO and limited available phase space with
long-wavelength excitations (analogous non-collinearity we also found in
CaFe$_{2}$As$_{2}$\cite{Sam}). These fluctuations can even destroy the long
range AFM order at lower R$_{Fe\text{-}Se}$ (or lower volume or $c/a$ ratio).
For larger R$_{Fe\text{-}Se}$ improving nesting conditions and a naturally
developing unstable longitudinal mode\ enhance the first effect: a stable,
ordered AFM state forms with large more localized $M$. The itineracy parameter
$\alpha$ coincides with the \textquotedblleft adiabaticity\textquotedblright%
\ parameter of Ref.\cite{LWA} and indicates that strongly frustrated metallic
systems simultaneously have strong non-adiabatic effects. Thus, our
calculations shown that transverse frustrations are rapidly decreasing as the
system becomes more localized.

To illustrate our result for FeSe we map the results obtained above onto
Heisenberg $J_{1}$-$J_{2}$-$J_{3}$ model\cite{Larkin}. Fig.4 clearly shows how
relatively weak exchange coupling between third NN transforms the collinear
ground state. Especially important are the third and forth NN coupling when
$J_{1}${$\approx$}2$J_{2}$, when the non-collinear state of different symmetry
can be stabilized by a very small coupling beyond second NN. In strongly
magnetic case (M=1$\mu_{B}$) when NN coupling appear to be very anisotropic
the ratio between main parameters of spin Hamiltonian are following (122
family): J$_{1y}$ $\approx30-40meV$, J$_{1x}\approx0.2$J$_{1y}$, J$_{2}%
\approx0.7$J$_{1y}$, J$_{z}\approx0.15$J$_{1y}$, K$_{1}\approx(0.2-0.3)$%
J$_{1a}$ and magnetic anisotropy $\sim$0.1 meV.

Despite theory predicting very similar properties for many families of iron
pnictides, experimental evidence for long range magnetic order in the normal
state have not been found in all compounds. In particular, no long range order
of any kind been found in LiFeAs or FeSe, nor have local moments been found so
far. In Tab.2 we summarize the calculated exchange coupling obtained for the
different iron pnictide systems, including some with no experimentally
established LRO. All these systems clearly demonstrate a stable AFM stripe
order at large moments. The exchange coupling shows rather itinerant behavior
for smaller moments (large range of parameters), and more localized for the
larger moment case. Due to large anisotropy between first NN exchange
parameters the striped structure is very stable and the isotropic $J_{1}%
$-$J_{2}$ Heisenberg model is totally not applicable for iron pnictide systems
at larger magnetic moments where this anisotropy is very large. However, this
model might be relevant at smaller moments when $J_{1}$ is close to $J_{2}$.
The similarity in the exchange coupling and the comparable stability of the
stripe order clearly contradicts the available experimental data for these
systems. Assuming that more experiments with better single crystals confirm
that LRO in LiFeAs and FeSe is absent, the fact that theory contradicts these
observations must be related to LSDA shortcomings. In particular, magnetic
zero-point fluctuations (absent in LSDA) may well be strong enough to suppress
the moment. Alternatively, there may be problems even in the LDA description
of bonding in iron pnictides at the independent-particle level.

Finally, we calculate the imaginary part of the bare dynamical transverse spin
susceptibility ${\chi^{0+-}\left(  \mathbf{q},\omega\right)  \ }$, which
provides insight into Stoner excitations\cite{MARKXI}. Calculations were
performed for the N{\'{e}}el structure at $z$=0.25, which corresponds to a
magnetic moment $M$=1.2$\mu_{B}$. Along $(110)$, ${\chi^{0+-}\left(
\mathbf{q},\omega\right)  \ }$ is particularly interesting; see Fig.5a.
%We focus on the low-energy range ($<200$~meV),
%which controls spin excitations and superconductivity.  Along $(110)$
Stoner excitations appear even for $\omega${$\rightarrow$}0 for small $q$. As
$q$ increases they disappear completely; see $\mathbf{q}=(0.25,0.25)$.
Finally, they reappear at still higher $q$.

Most of these novel features can be explained in terms of the unique nature of
the Fermi surface. The Fermi surfaces of FeSe (and the Fe pnictides) are
approximately cylindrical, and are centered either at $\Gamma$ or at the
($\pi,\pi$) point in the $xy$ plane, as shown in Fig.3 of Ref.\cite{SinghFeSe}
Cylinders at $\Gamma$ are of hole character; those at the zone corner are of
electron character. $\mathbf{q}$ is the connecting vector that links an
electron at $\mathbf{k}+\mathbf{q}$ to a hole at $\mathbf{k}$. ${ \chi^{0+-}
\left(  \mathbf{q}, \omega\right)  \ }$ is obtained from a sum over all
allowed $\mathbf{k}$ (i.e. any $\mathbf{k}$ for which an electron state at
$\mathbf{k}+\mathbf{q}$ and a hole state at $\mathbf{k}$ are separated by an
energy difference $\omega$). We have identified the following facts, which can
explain the behavior in Fig.5a.

$(a)$ For $\mathbf{q} <(0.25, 0.25, 0 )$, $\mathbf{k}$ points only couple to
points $\mathbf{k}+\mathbf{q}$ in the \textit{same} pocket.

$(b)$ For $\mathbf{q} >(0.25, 0.25, 0 )$, $\mathbf{k}$ points only couple to
points $\mathbf{k}+\mathbf{q}$ in \textit{other} pocket.

$(c)$ For $\mathbf{q}=(0.25, 0.25, 0 )$, no two $\mathbf{k}$ points can be connected.\ 

$(d)$ For $\mathbf{q} >(0.375, 0.375, 0 )$. there is a strong inter-pocket excitation.

Along (100) the story is different. Since the two pockets are separated by
$(1/2,1/2,0)$, they can not communicate for $\mathbf{q}$=$(q00)$ (at least at
low energy). So only intra-pocket excitations, $\Gamma${}$-${}$\Gamma$, M$-$M,
are allowed. This picture is borne out in Im${ \chi^{0+-} \left(  \mathbf{q},
\omega\right)  \ }$, shown in Fig.5b. Im${ \chi^{0+-} \left(  \mathbf{q},
\omega\right)  \ }$ is already present for small $\omega$, e.g. 50~meV, when
$q$ is small; but almost disappears when $q$ reaches 0.25, and completely
disappears for larger $q$.

While we have not yet attempted to calculate spin fluctuations, the fact that
large Stoner excitations are present in a significant fraction of the
Brillouin zone offer strong indications that low energy spin-flip excitations
are important. They can strongly affect the single-particle description of the
magnetism; in particular they reduce the magnetic moment $M$. These
fluctuations may explain the discrepancy between observed magnetic order, and
that predicted by the LDA in case of FeSe.

In conclusion, two competing mechanisms in the newly discovered
superconductors have been identified: strong geometric frustrations dominant
for small moments, which allow distant interactions lead to a symmetry
breaking, forming a continuum of non-collinear states. As $M$ increases,
magnetic Jahn-Teller effect\cite{Zaa} creates better nesting conditions with
stabilzation of an AFM striped configuration. We have found third and fourth
neighbor couplings $J_{3}$ and $J_{4}$ to be the driving force for
non-collinear spin fluctuations. We believe that is the generic feature of
iron pnictides when the Fe moment is small. The calculated exchange coupling
and the corresponding spin wave excitations along z-direction are relatively
small, can be easily destroyed by temperature and two-dimensional character of
magnetism is likely to be realized at higher temperatures or paramagnetic
state. We also find that strong frustration --- a condition necessary for
excess entropy in order to admit a superconducting state --- are favored in
the itinerant magnets when conditions for adiabaticity and long wavelength
approximation are not well satisfied. Here longitudinal fluctuations make an
additional contribution to the spin entropy. As moment increases, the in-plane
exchange coupling along 'ferromagnetic' line changes its sign, so the stripe
order becomes absolutely stable for such 'no frustration' localized magnetic
moment case. We identified those magnetic excitations in pnictides that may
exist below Neel temperature and contribute to observable properties. Together
with traditional spin waves, in this system coexist Stoner type of excitations
(electron-hole pairs) and longitudinal spin fluctuations, especially at small
moment. The presence of low energy Stoner excitations ultimately identifies
this system as itinerant magnetic system. In addition to such traditional
excitations, this system has very peculiar four fold symmetry planar spin
fluctuations with the energy of 15 meV and below (modes (b)-(d) on Fig.1) with
some of them being similar to 'nematic' spin order. Our results indicate that
the localized spin description is somewhat academic for this system if
magnetic moment does not exceed 1$\mu_{B}$. Overall, the magnetic
transformation in iron pnictides can be described as a transition from
frustrated itinerant moment phase to symmetry broken orbitally ordered
localized magnetic moment phase with "striped" order. The transition happened
around $M$=$1\mu_{B}.$

Since this paper has been submitted new neutron experimental data have been
published for FeTe$_{1-x}$Se$_{x}$\cite{FETESE}. The authors observed the
presence of two-dimensional incommensurate excitations which may reflect the
non-collinear ordering identified above.

Work at the Ames Laboratory was supported by Department of Energy-Basic Energy
Sciences, under Contract No. DE-AC02-07CH11358. ASU was supported by DOE
contract DE-FG02-06ER46302 and ONR grant N00014-07-1-0479, and is grateful to
the Fulton High-Performance Computing Center for use of their facilities.

\begin{tabular}
[c]{|l|l|l|l|l|l|}\hline
& $n$ & \ \ R & $\overrightarrow{\tau}$ & $J_{T}$ (mRy) & $J_{L}$
(mRy)\\\hline
$J_{100}$ & 2 ($\uparrow$) & 0.497524 & $0.5,0,0$ & -0.508 & -0.644\\\hline
$J_{010}$ & 2 ($\downarrow$) & 0.50 & $0,0.5,0$ & 0.773 & 0.828\\\hline
$J_{110}$ & 4 ($\downarrow$) & 0.705358 & $0.5,0.5,0$ & 0.145 & 0.299\\\hline
$J_{200}$ & 2 ($\uparrow$) & 0.995 & $1.,0,0$ & 0.045 & -0.045\\\hline
$J_{020}$ & 2 ($\uparrow$) & 1.0 & $0,1.,0$ & 0.046 & -0.043\\\hline
$J_{001}$ & 2 ($\uparrow$) & 1.028 & $0,0,1$ & 0.001 & -0.008\\\hline
$J_{120}$ & 4 ($\uparrow$) & 1.1169 & $0.5,1.,0$ & 0.060 & -0.099\\\hline
$J_{210}$ & 4 ($\downarrow$) & 1.1136 & $1,0.5,0$ & -0.053 & -0.013\\\hline
\end{tabular}

Table.1 The first eight Fe-Fe transverse $J_{T}$, and longitudinal $J_{L}$
exchange parameters in FeSe. Column $n$ denotes the number of equivalent
nearest neighbors, together with the spin orientation. R denotes the distance
to the central atom and $\overrightarrow{\tau}$ the connecting vector in units
of the lattice parameter $a$. $J_{T}$ is related to the Heisenberg model $J$
as $J\mathbf{S}_{i}\mathbf{S}_{i}=J_{T}$ and can be defined as a second
derivative of the LSDA total energy with respect to moment variation.%

\begin{tabular}
[c]{|l|l|l|l|l|l|l|l|l|l|l|l|l|}\hline
& M & $J_{100}$ & $J_{010}$ & $J_{110}$ & $J_{200}$ & $J_{020}$ & $J_{120}$ &
$J_{210}$ & $J_{001}$ & $J_{101}$ & $J_{011}$ & $J_{0}$\\\hline
CaFe$_{2}$As$_{2}$ & 1.61 & -1.5 & -16.1 & -12.1 & -2.0 & -5.0 & -3.3 & -0.3 &
-0.8 & -2.0 & -0.4 & 91.5\\\hline
SrFeAsF & 1.78 & -8.9 & -41.8 & -7.8 & -0.5 & -0.5 & 3.7 & -0.2 & 0. & 0. &
0. & 102.6\\\hline
LiFeAs & 1.28 & -11.2 & -35.5 & -7.6 & -3.2 & -3.1 & 3.6 & -0.5 & -0.8 &
-0.3 & -0.5 & 91.2\\\hline
FeSe & 1.07 & -45.3 & -70.9 & -13.5 & 0.8 & -0.1 & 4.5 & -2.2 & -0.3 & -0.2 &
-0.1 & 35.9\\\hline
\end{tabular}

Table.2 The several Fe-Fe Heisenberg model parameters $J$ \ in different
families of iron pnictides (in meV). These parameters have been obtained from
theoretically calculated $J_{T}$ by normalizing it by scalar product
$\mathbf{S}_{i}\mathbf{S}_{j}$ where $\mathbf{S}_{i}$=$\mathbf{M}_{i}$/2.

Fig.1 The dependence of magnetic moment on Fe site as a function Fe-Se
distance (R$_{Fe-Se}$) for the striped (lowest energy) and simple N{\'{e}}el
(highest energy) magnetic structures. Also shown are schematics of the
magnetic structures considered in the calculations: (a) striped structure (b)
"zig-zag" structure. A transition from a striped to "zig-zag" structure (a
'nematic' type of state) leaves AFM ordering between first NN atoms unchanged
(j$_{1}$ mode). (c) "rotator" structure (or so called 2$\mathbf{k}$
structure). Here AFM ordering between second NN is unchanged (j$_{2}$ mode)
(d) mixed structure with AFM moments along diagonal and non magnetic NN atoms
(e) dimer structure (f) simple N{\'{e}}el structure.

Fig.2 Magnetic interaction parameters in FeSe for the striped structure, as a
function of Fe magnetic moment $M$ (controlled by varying R$_{Fe-Se}$).
$J_{1x}=J_{100}$ and $J_{1y}=J_{010}$ are the NN interaction along $x$ and $y$
respectively; $J_{2}=J_{110}$ is the second-NN interaction in the plane
(vectors are shown in the Table.) Interactions between planes are small
($\sim6$\% of $J_{2}$). Insert shows the anisotropy of the NN exchange
coupling as a function of Fe moment. The inversion of the sign of the exchange
coupling paramer $J_{1x}$ occurs at M$\approx$1.88 $\mu_{B}.$

Fig.3 The total energy of a spin spiral as a function of spiral wave vector
$q$ near the AFM striped structure. Top: SS for \textbf{{q}}=(0,$q$,0)2$\pi
/a$. Near $q$=0.25 (not shown), the moment disappears entirely. Bottom: SS
along the stripe, for \textbf{{q}}=($q$,0,0)2$\pi/a$. The minimum energy
occurs at $q\approx{}0.02$, corresponding to a slow spiral along "frustrated"
direction $x$.

Fig.4 The phase diagram of $J_{1}$-$J_{2}$-$J_{3}$ model and calculated
non-magnetic susceptibilities ratio for FeSe (central red star). Other stars
correspond to 2\% and 4\% of hole (right) and electron (left) doping of FeSe.
In all cases the finite value of $J_{3}$ coupling leads to the stabilization
of non-collinear order of different symmetry. The relative strength of $J_{3}$
is increased as moment decreases and the system becomes more itinerant.

Fig.5 Bare transverse spin susceptibility ${ \chi^{0+-} \left(  \mathbf{q},
\omega\right)  \ }$ as a function of $\omega$ for different $q$ points along
$(100)$ and $(110)$.

\bigskip\bigskip\bigskip


\begin{thebibliography}{99}
\bibitem {NOTE}Since the LDA+U method selects by hand the on-site Hund energy,
applying it in this context is problematic. On the other hand, the LDA
successfully predicts important properties in the Fe pnictides, such as the
Fermi surface geometry and the prediction of a striped magnetic ground state.
\end{thebibliography}
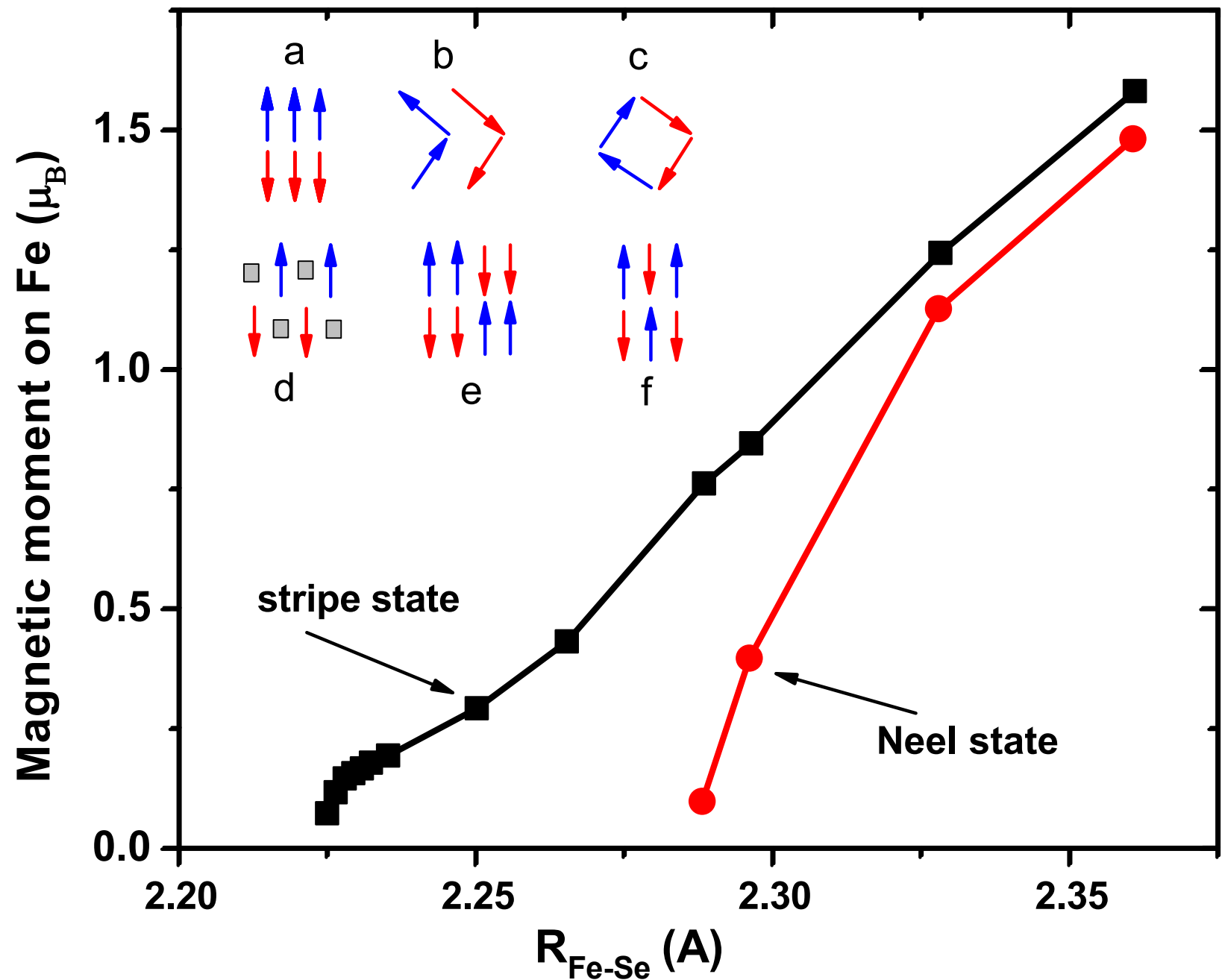
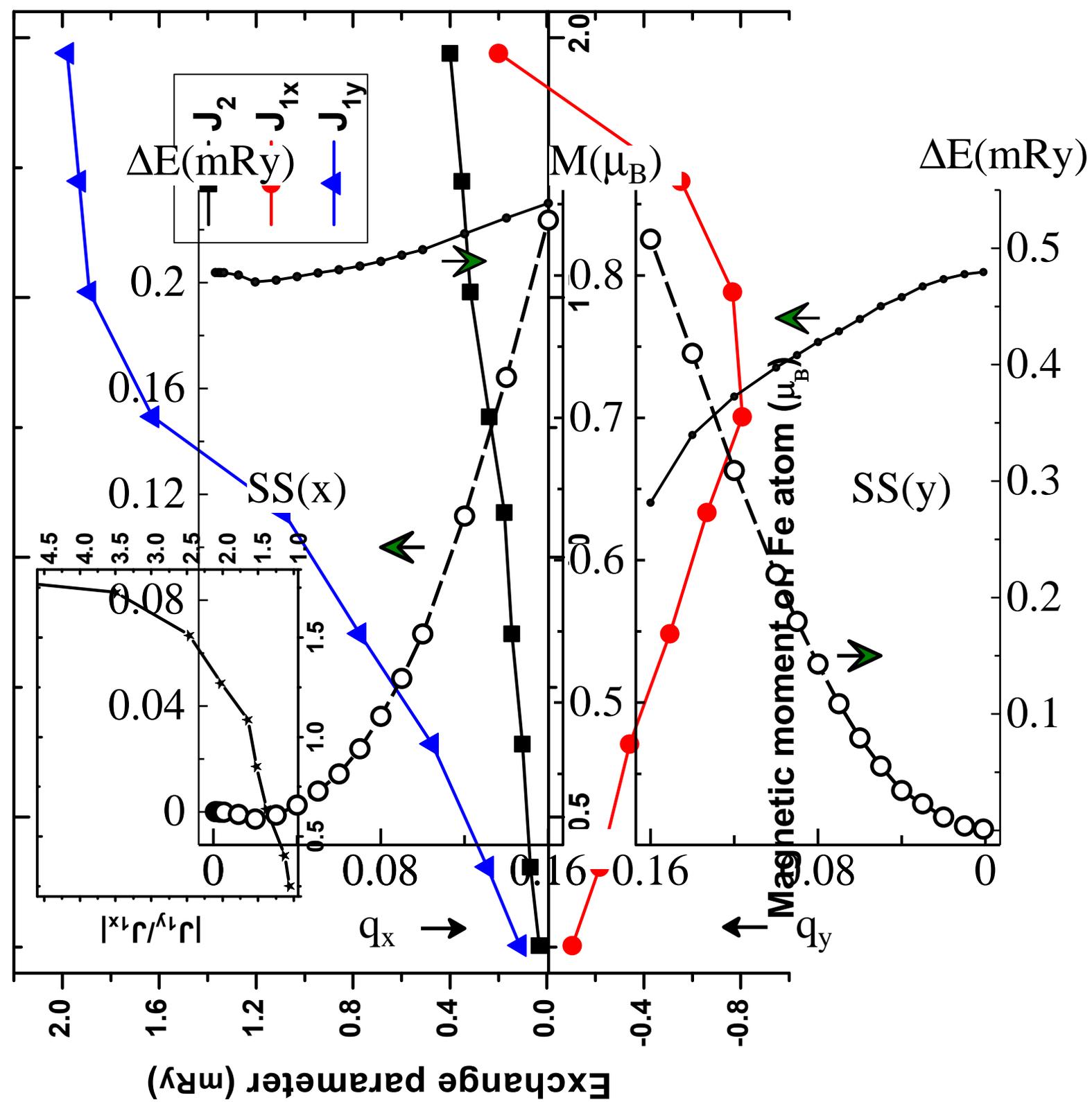
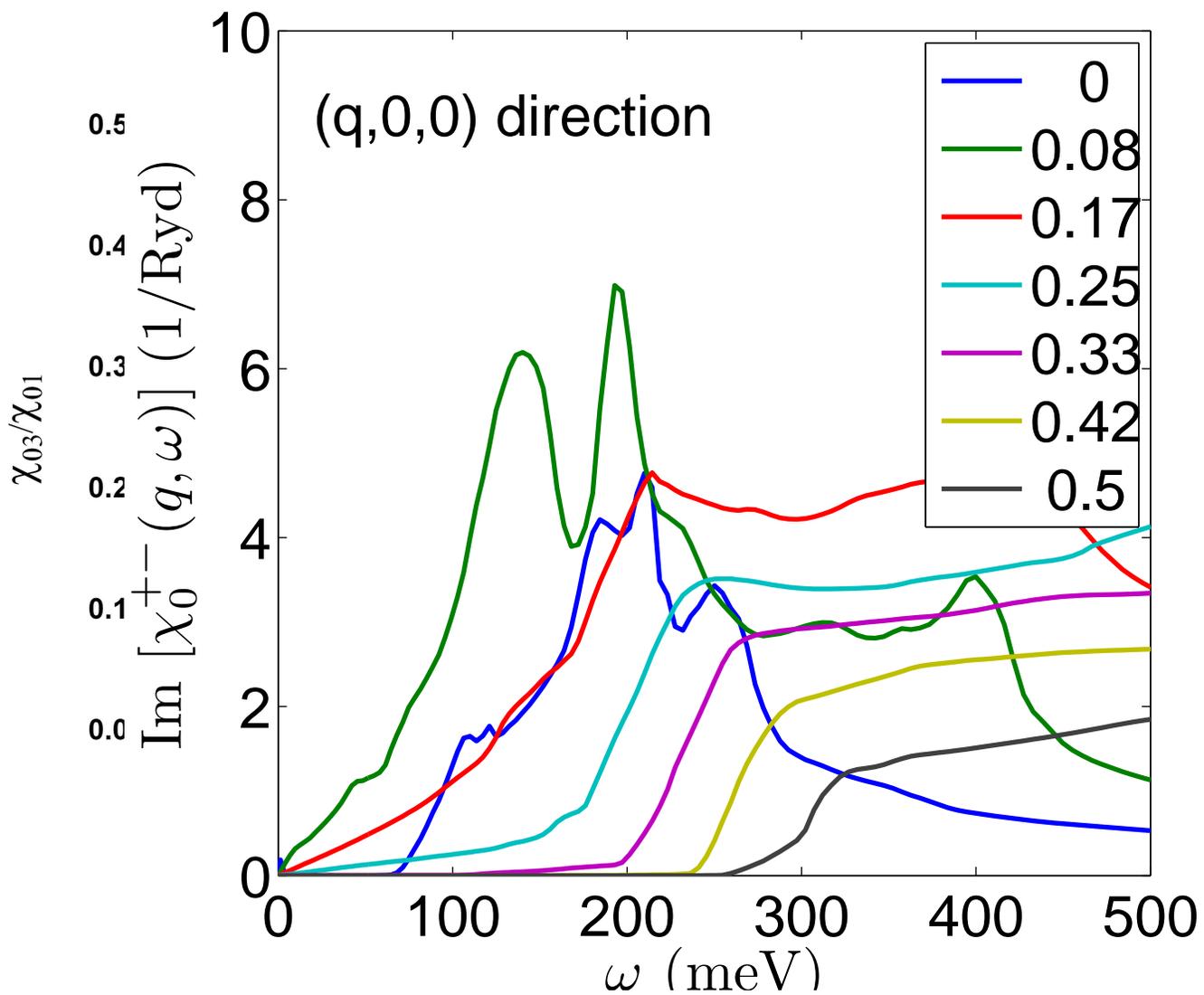
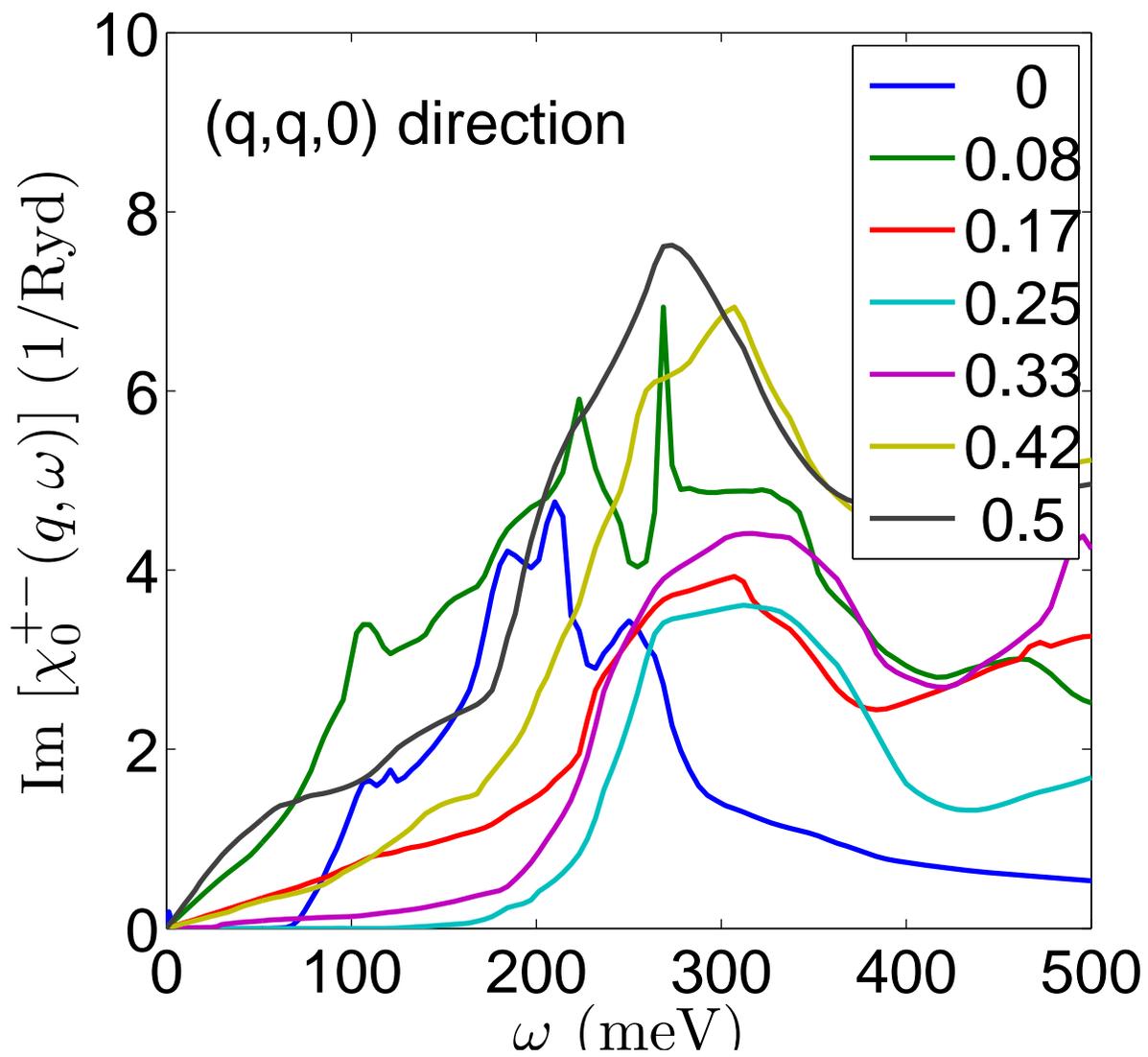
\end{document}